# Simple and Complex Metafluids and Metastructures with Sharp Spectral Features in a Broad Extinction Spectrum: Particle-Particle Interactions and Testing the Limits of the Beer-Lambert Law

Lucas V. Besteiro[1*], Kivanc Gungor[2], Hilmi Volkan Demir[2,3], Alexander O. Govorov[1,4*]

[1] Department of Physics and Astronomy, Ohio University, Athens, Ohio 45701, United States

[2] Department of Electrical and Electronics Engineering, Department of Physics, UNAM-Institute of Materials Science and Nanotechnology, Bilkent University, Ankara 06800, Turkey

[3] Luminous! Center of Excellence for Semiconductor Lighting and Displays, School of Electrical and Electronic Engineering, Division of Physics and Applied Physics, School of Physical and Mathematical Sciences Nanyang Technological University, Singapore 637371, Singapore

[4] Institute of Fundamental and Frontier Sciences, University of Electronic Science and Technology of China, Chengdu 610054, People's Republic of China

## Abstract

Metallic nanocrystals (NCs) are useful instruments for light manipulation around the visible spectrum. As their plasmonic resonances depend heavily on the NC geometry, modern fabrication techniques afford a great degree of control over their optical responses. We take advantage of this fact to create optical filters in the visible-near IR. Our systems show an extinction spectrum that covers a wide range of wavelengths (UV to mid-IR), while featuring a narrow transparency band around a wavelength of choice. We achieve this by carefully selecting the geometries of a collection of NCs with narrow resonances that cover densely the spectrum from UV to mid-IR except for the frequencies targeted for transmission. This fundamental design can be executed in different kinds of systems, including a solution of colloidal metal NCs (metafluids), a structured planar metasurface or a combination of both. Along with the theory, we report experimental results, showing metasurface realizations of the system, and we discuss the strengths and weaknesses of these different approaches, paying particular attention to particle-particle interaction and to what extent it hinders the intended objective by shifting and modifying the profile of the planned resonances through the hybridization of their plasmonic modes. We have found that the Beer-Lambert law is very robust overall and is violated only upon aggregation or in configurations with nearly-touching NCs. This striking property favors the creation of metafluids with a narrow transparency window, which are investigated here.

# Introduction

Plasmonic materials, such as noble metals, afford us the capability to control light in ways and at wavelengths that other materials cannot. There is a growing body of work studying multi-particle or continuous material behavior in the nanoscale to create exotic optical phenomena, such as negative-index,[1,2] gradient-index metamaterials,[3] and flat optics.[4] Furthermore, plasmonic nanocrystals show strong optical resonances and admit a great degree of control in tuning their frequency, extending into visible and infrared, by tailoring their geometry.[5-7] This has opened up a variety of possibilities to create novel optical responses with plasmonic NCs.[6,8] A phenomenon of interest, discussed in this paper, consists in creating extinction features that allow for frequency selection on the incident light. We will describe the optical properties of composite materials with a narrow transparency window inscribed into a broad extinction profile, using collections of nanoparticles and metastructures with plasmonic resonances. Such collections should have isotropic optical properties and will appear in a range of packing conditions, from sparse to dense. In metafluids with tightly-packed elements, the near-field inter-component interactions become moderate or strong and can impact the performance of the system. This is in contrast to our previous studies of NC collections with small packing densities, which we called "simple" metafluids.[9,10] In the conceptual model of simple metafluid, a single ballistic photon, propagating through the system, interacts with a large collection of specifically designed nanocrystals. This system of NCs is diluted and the NCs in such system do not interact strongly via the plasmonic near-fields. However, a coherent light wave does couple with all NCs as it directionally propagates through the solution. The transmission through a simple metafluid is given by the Beer-Lambert law.

Here we focus on the effects of the particle-particle interactions via both near- and far-fields induced by NCs. The density of NCs is now treated as a varying parameter. In this paper, our models for metafluids consist of densely packed NCs, and we regard such systems as complex metafluids. Correspondingly, the Beer-Lambert law is broken at some critical packing densities of NCs, and it is interesting to investigate such critical densities. Overall, our models show that the Beer-Lambert law is remarkably robust. This agrees with the wide applicability of this law, be it qualitatively or quantitatively, on so many experimental systems, including so-called muddy media with strong internal scattering.

In our study, we consider two types of experimental systems: a solution of colloidal NCs and a 2D structured metastructure. In the metafluids considered here, controlling the exact composition of building blocks makes it possible to design very unusual isotropic transmission spectra with sharp optical features such as a transmission window or a stop band. In the case of using metastructures as building blocks, an exact organization of its layout is another crucial factor to achieve our goals. Optical features such as a transmission window or a stop band are in general not possible or difficult to achieve with molecular compounds, such as dyes, as they provide narrow absorption lines within a relatively small range of frequencies, typically in the visible spectrum.

Our proposal exists among other ways to approach this objective: some exploit properties of plasmonic materials by creating highly structured plasmonic multilayer metamaterials,[11,12] or carefully arranging individual nanoparticles such that their interacting plasmonic resonances exhibit plasmonic Fano effect,[13-17] which involves the destructive interaction of a broad excited state with a narrow resonance. Results obtained with these approaches offer very strong and sharp transparency spectral features, but they are generally strongly angle-dependent and sensitive to relative NC positions.

The approach that we describe and discuss in this article, originally presented in our previous work in refs 9 and 10, aims for a system with isotropic transmission spectrum. To our knowledge, our proposal is unique in so far as (a) it offers extinction over a large range of frequencies surrounding the transparency window, (b) our system with sharp optical features is also optically isotropic, and (c) different wavelengths can be targeted to be used as window without changing the fundamental methodology. The common idea behind the systems reported here is the use of narrow single crystal plasmonic resonances to densely cover a broad range of wavelengths, while leaving exposed a well-defined narrow range that will form the transparency window. Others have pursued a similar goal by ablating the system at a selected transparency frequency,[18] but our approach involves an a priori design process in which we select a collection of NCs that provide suitable resonant modes. The selection process is made accessible by numerical methods performed with computational aid, which allow for a careful modeling process that helps to identify geometries of interest.[9,10] Of course, this is necessarily paired with modern fabrication methods that allow control over the fabricated NC shapes.[19-21]

Our choice for the numerical approach involves classical electrodynamic equations solved self-consistently with Finite Element Methods. As other numerical methods which solve a classical system in a volumetrically defined system, it has been shown to be adequate to study arbitrary plasmonic geometries, and it includes information about retardation effects. These formalisms, however, are certainly not alone in providing a good description of plasmonic systems, and novel approaches to the computational study of plasmonic complexes are present in the literature, showing the capability to efficiently model systems of great complexity.[22]

For a simple metafluid, our fundamental assumption of the modeling process is the treatment of the NCs as being in a regime of weak near-field interaction, where the plasmonic resonances are only broadened by intrinsic material properties influencing plasmon dephasing processes, not by interparticle hybridization due to electromagnetic interactions. Under this assumption, one can model the resulting global optical filter as a linear superposition of the extinction cross-sections of individual NCs. Then, since the model allows us to treat NCs in a simple metafluid as independent components, we can, of course, construct very advanced and tailored optical spectra with sharp features. However, both the approximation and the effect break apart in systems where the NCs are close enough to interact strongly with their neighbors via plasmonic near-fields. This paper illustrates this fundamental behavior using two realistic cases in which the Beer-Lambert law does not hold. These cases involve strongly interacting NCs and sharp optical features, such as a transmission window, are either destroyed or strongly distorted.

Together with an expanded theoretical part, we include an original experiment with a plasmonic metastructure on a flat surface. An exact organization of the individual

NCs in such metastructures, which can potentially be used as a metaflake, assures the survival of the transmission window effect even for a tightly-packed geometry in which the NC-NC interactions are strong. Another model in our study is a metafluid based on the experimental system from ref 10. In this case, the near-field interactions in a NC solution arise from NC aggregation that can, of course, destroy the sharp spectra.

This paper will be organized in the following way. The first section describes the theoretical methodology employed in the rest of the paper. Then, we will cover colloidal NCs solutions, planar structured metasurfaces and a possible combination of the previous two approaches in a semi-structured metaflake solution. Experimental results are reported for the first two families of systems.

## 1. Theoretical approaches

The material feature that we will mainly discuss, as it is central to the description of the effect that we want to study, is its transmittance:

$$T = \frac{I_t}{I_i} = 10^{-OD} \quad (1)$$

which is the fraction of light intensity incident on the sample that does pass through it (Figure 1a). The optical density $OD$ is a derived magnitude related to transmission by the expression above. It is particularly useful to model a material in terms of its optical density when we can make use of the Beer-Lambert law. This law relates the optical density of a solution with the concentration of the solute, of which we have the means to calculate its extinction cross-section, and the optical path. Such relationship reads

$$OD = \frac{L}{\ln(10)} \sum_i \sigma_i n_i \quad (2)$$

where $L$ is the optical path traversed by light in the material. The sum index runs through the different NCs in the solution, $\sigma_i$ and $n_i$ denoting the extinction cross-section and number concentration, respectively, of the $i$-th species. In the systems that we will describe these species will be the different geometries of NCs that comprise the full dissolved or constructed ensemble. Since individual NCs in a metafluid are generally anisotropic, the optical parameter $\sigma_i$ used in eq 2 should be calculated as an excitation cross section averaged over all directions and polarizations of incident light.

This commonly used theoretical model (eqs 1 and 2) can be understood as an effective-medium approach to the problem of a coherent light beam propagating through a medium with randomly-dispersed absorbers and scatterers. The optical density (2) in the Beer-Lambert law is, in fact, an optical parameter describing an effective medium. Another related theoretical approach is the Maxwell-Garnett model that describes an effective dielectric function of a composite medium in the limit of small sizes of inclusions.[23] Both the Beer-Lambert law and the Maxwell-Garnett model stand as key effective-media approaches in optics of composite media.[23]

**Electromagnetic simulations.** The electrodynamic calculations behind the theoretical results shown here have been performed within a classical framework, solving the Maxwell equations using Finite Elements Methods. Particularly, we have used the commercial package COMSOL. Cross-sections obtained for the system in solution come from considering non-periodic systems in where we average the contribution from six different illumination conditions (along the three main axes, with two orthogonal linear polarizations). These cross-sections are then used in combination with the Beer-Lambert law. This approach is also used for the metaflakes solution. The planar metasurfaces are in turn calculated using periodic boundary conditions and considering only illumination with linear polarizations orthogonal to the surface. In this paper we have used Au NCs, which we have modeled with bulk experimental permittivity data.[24] Further details about the material data used in our simulations can be found in the Supplementary Information.

**The Beer-Lambert law and a high concentration regime**. Beer-Lambert law has some well-known limitations. Presuming a homogeneous solution, it provides a good description of the light transmission only for relatively low concentrations of the suspended particles.[25] In general, this law does not hold for high solute concentrations. This is so because by increasing concentration the average distance between solute components will decrease, which in turn implies that an increasing proportion of them would interact with their neighbors. In such scenario, the simple linear relationship described in eq 2 does not hold for the entire NC population.

It is then clear that, to model light transmission in systems with non-negligible interparticle interaction, one should either do without the Beer-Lambert law, or describe the systems in some modified fashion. One possible way to preserve the very convenient expression in eq 2 is to cluster or group NCs into independent cells. Simultaneously, one should assume that the total NC density in solution remains low. In other words, this approach assumes that the total NC density in solution is low, but the local density of NCs in clusters or metastructures can be high (Figures 2, 3 and 8). Then, one can calculate the cross-section profile of the cluster that includes information about their interparticle interaction. The group of particles would be then considered a species of its own, with a joint extinction cross-section and an appropriately modified concentration weighing it. We then obtain the means to connect interacting NCs extinction cross-sections to the convenience of eq 2 and describe physical systems with a high local concentration of NCs or in conditions when the NCs aggregate.

Mathematically, the essence of the Beer-Lambert law is a linear superposition of optical cross sections (eq 2) that assumes weakly-interacting species. When we group our NCs into complexes or clusters, we can again write down a similar law of propagation in terms of the new cross sections:

$$OD_{metafluid} = \frac{L}{\ln(10)} \sum_{\alpha} \sigma_{\alpha} n_{\alpha} \quad (3)$$

where $\sigma_{\alpha}$ and $n_{\alpha}$ are the cross sections and concentrations, respectively, for the new complexes. Again, the system of NC complexes should be assumed to be diluted, but, within a complex, NCs can be strongly packed. Therefore, the cross sections of complexes ($\sigma_{\alpha}$) can be strongly changed as compared to the sum of the original ones ($\sigma_i$), owing to strong near-field plasmonic interactions. When such interactions become very strong, interesting and narrow optical features, such as a transparency window, can vanish. We will see such behavior in the following sections.

**Strength and character of the NC-NC interactions.** In some physical situations, it is very convenient to treat inter-particle coupling using the dipolar approach.[26] We should note that ours is not such a case. Our aggregated systems of NCs and metastructures are often not in the dipole regime of interaction and, therefore, exact electromagnetic simulations are needed. Two particular examples, discussed below, are aggregated parallel nanorods and meta-flakes in which the inter-particle distances are smaller or comparable with the NC dimensions. Therefore, our approach has an effective multi-dimensional character: Simultaneously with the exact treatment of individual complexes at small scales, we still use an approximation of effective optical media (Beer-Lambert law) at much longer scales.

## 2. Application to NC solution: A simple metafluid

Within the simple framework described above, we approach the goal of designing an isotropic EM filter that provides a narrow ($\sim 100\, nm$) transmission window into the visible to near infrared spectrum, surrounded by a broad extinction band. This is done by carefully selecting the ensemble of gold NCs to be deployed in aqueous solution. The solvent is also instrumental in determining the optical characteristics of the solution. As water absorbs light in the IR, the effective extinction region will be wider than reported, extending to the region $\lambda > 1200\, nm$. Additionally, regions of the spectrum far from the transparency window can also be padded with dielectric NCs, as shown in ref 10. We can therefore reach the intended goal with a relatively low variety of NC geometries (as compared, for instance, to the metasurface in a following section of this paper), because the spectral region that we need to cover with the NCs is narrower. In other words, what the solute achieves is to reduce the width of the transmission window of the solution by one order of magnitude, from $\sim 1\, \mu m$ to $\sim 100\, nm$, in comparison with the solvent alone and, crucially, it does it in a targeted manner to specific frequencies.

Our proposed example design involves four different geometries of cylindrical Au rods immersed in a matrix with $\varepsilon_0 = 1.77$ (Figure 1b). Their sizes, expressed as (diameter)x(length) in nm, are 25x36, 10x45, 10x58 and 10x81, respectively, in accordance with the experimental sizes in ref 10. Shapes are selected so their plasmonic resonances cover a relatively broad band of wavelengths, while their interaction with light is weak inside the intended transparency window, around $\lambda = 740\, nm$. We used a nanorod shape for the NCs because of the strong and narrow plasmonic resonance that it affords, making it adequate to target relatively specific wavelength ranges. In our calculations, we have assumed that each component (one kind of NCs) is monodisperse. Although it would be interesting to extend the study to the impact of polydispersity over the window effect, the fact that this effect is observed in the experimental realization of the metafluid,[10] where samples cannot be assumed to be perfectly monodisperse, suggests that size heterogeneity of experimental Au nanorods of high quality does not destroy the window effect. On the other hand, in our simulations we effectively introduce some degree of polydispersity by using the broadened dielectric function (See Supporting Information). Then, the optical window got broadened in Figure 2d, but the transparency effect in general survived.

With an already defined selection of NCs, we can calculate the expected transmission for samples with different concentrations under the assumption of no interparticle interaction, but also study different scenarios where this hypothesis does not hold. However, given the large number of particles involved, one of course expects that some proportion of the NCs would be interacting at least weakly with some other neighboring gold NCs, and even that a subset of them would be aggregated with others. In other words, if we consider a solution as a collection of independent volume units with a certain number of NCs in it, we expect that these unit volumes to show a gradation of cases from no NC-NC interaction to full aggregation. In Figure 2, panels b through e, we explore a small fraction of simple cases that could be encountered in this progression. The different transmission spectra correspond to different individual unit volume configurations, but maintaining the same global number density of the solution ($n = 1.1 \cdot 10^{11} \ cm^{-3}$, also assuming an optical path of $L = 1 \ cm$), and always considering the same NR-shape proportion in each unit volume (two instances of each rod size).

**Experimental details for the nanorod metafluids**. The experiments used here to illustrate the concept of complex metafluid were reported in ref 10. In that work, the solution exhibiting transparency windows were fabricated by mixing a B-doped Si NC colloidal solution predispersed in water with a mixture solution of CTAB-capped Au NR. Four different samples of Au NRs were combined, with dimensions of 25 nm x 36 nm, 10 nm x 45 nm, 10 nm x 58 nm and 10 nm x 81 nm, respectively. Transmission measurements were performed using a spectrophotometer, with the solution loaded in a quartz cuvette with a 1 cm light path. Transmittance spectra of the empty cuvette was used as reference. Particle aggregation was increased by extensive centrifugation of the sample to remove some of the CTAB ligands stabilizing the Au NRs. Further details can be found in ref 10.

## 3. Pair-wise interaction in the solution: A complex metafluid

The transparency profiles shown in Figure 2d suggest that, given a suitable concentration of NCs and an appropriate size selection, a narrow transmission window effect could be achieved in a system with negligible NC-NC interaction. As we step into configurations that involve some degree of NC interaction we observe both displacement and an increased opacity in the transmission window. This is as one would expect, because NC-NC near-field interaction does effectively broaden, displace and overall distorts the individual NC resonances. It is remarkable, however, the relatively large degree of aggregation required to completely break the transparency window. Particularly, this effect is for the most part resistant to pair-wise NC interaction. In this scenario there is no single non-interacting gold NC and the rods in every pair are arranged parallel to each other at close distance (2 nm), so that their resonant dipolar modes are in interaction, yet the phenomenon is robust enough to withstand this non-trivial distortion of the devised plasmonic resonances.

The transparency window stability is more strongly impacted, though, if we consider samples in where we include near-field interaction among more than two rods. Particularly, Figure 2d shows results for three configurations with different relative distances and orientations between the rods, depicted in panel b of the same figure. The first two systems show NCs arranged in a disordered fashion. The multiple NC-NC interactions are enough to greatly reduce and shift the window, more so when the particles are brought to close proximity.

But the most disadvantageous configurations would be the ones in which the rods are closely packed and collinear. A representative instance of such situation is also included in Figure 2, where not only the transmission window disappears, but the broad extinction coverage is also prevented by the rods' mutual interaction. In this configuration we align all the NC dipole resonances in parallel. In one perspective, we could say that this leads to the multiple plasmonic excitations to hybridize jointly; alternatively, we can also consider this as blocking the possibility of strong dipolar responses at the intended wavelengths, as the dynamic charge screening is now performed by an effectively thicker material structure. This disrupts the once organized set of individual narrow resonances, breaking down not only the afforded transparency window, but also the broadband extinction coverage of the NC sample.

This qualitative progression from non-interacting rods to intentionally packed groups is illustrated in Figure 2e. The calculated transmission values at the center of the nominal transparency window are there arranged from sparse to packed. This figure provides grounds for this simple logic to deal with aggregation (although we will problematize it below with some additional nuance). Furthermore, it offers a good qualitative understanding of the experimental observation, yielding good quantitative agreement as well. Interestingly, this is done while retaining the simple and useful theoretical perspective of the Beer-Lambert law.

Regarding details of modeling in Figure 2, we assumed a constant cell density, $n = 1.1 \cdot 10^{11}\ cm^{-3}$. Then, the total concertation of NCs should be counted as $8 \cdot n$ since, in our approach, one "elementary set" of NRs contains 8 NCs. In the case of paired nanorods, we can count 10 different configurations of a two-NC complex (Figure 2b) which have the same probabilities of formation and correspondingly the same concentrations. In the case of aggregates, we simply aggregate 8 NCs.

Relevant experimental results are shown in Figure 2f (system with same NC distribution as described here, reproduced from ref 10). By comparing these experimental results with the theoretical data, one could read into their similarities that the average NC into the real solution is interacting only weakly with others. We believe that this validates the simple, yet predictively powerful, design approach that we describe here.

However, before continuing the discussion with the planar metasurface, we can observe in more detail the relationship between NC-NC interaction and aggregation. In Figure 3 we show an additional perspective of the system's evolution towards the solute aggregation. Against the benchmark of a perfectly diluted collection of NCs, we present averaged isotropic results for the extinction cross-section of NC sets inside a decreasing unit volume (Figure 3a). From that data, Figure 3c is created by calculating the transmission of a system with such extinction profile. Data points at $\lambda = 740\ nm$ are highlighted in Figure 3d, with the local density on the horizontal axis ($n_{local} = V_{complex}^{-1}$), computed as the inverse of the minimal rectangular volume that contains each cluster of NCs, and working as a metric for aggregation. A quick observation on these data reveals a clear inverse relationship between the density of the individual cell and the transmission at the center of the nominal transparency window, as expected both from what we have seen in Figure 2 and from a simple theoretical perspective. On closer inspection, however, we can see that this relationship is not as simple as we assumed in the above discussion. We think that two characteristics of the plot illustrate this point: (1) the least aggregated collection offers a transmission that is higher than the one for the ideal system and (2) small variations in compactness can have large variations in opacity (configurations 2 to 4, for instance). These out-of-tendency variations arise from the fact that the relative orientations of the NCs involved have a critical influence over the changes in the optical response of the system. Critically, NRs can be relatively oriented in such a way that extinction cross section for the complex decreases at the window's center. It is clear that a fine-grained description of the solute configurations has to show this level of sensitivity to spatial arrangement, but it is also evident that it will be confounded in the macroscopic results, where remains a general tendency towards the distortion of the transparency window with aggregation. To summarize this discussion, we also note that the arrow in Figure 3d shows just a qualitative tendency related to the aggregation and any quantitative measure of the effect of aggregation on the optical spectra would require sampling of many more NC complexes.

Of course, for this theoretical description of the interaction between colloidal NCs to better resemble the actual physical system, we could consider the full statistical ensemble over relative positions and orientations of NCs. It is clear, however, that to do so while still relying on full electrodynamic solutions to measure particle interaction is not only impossible, but unnecessarily over-descriptive. We are confident that our selected examples are enough to show, not only the relevance of the general trend concentration–interaction discussed, but also the relative robustness of the transparency effect, particularly upon the favorable comparison with experiment, and thus the adequacy of the fundamental design approach proposed.

## 4. Metastructures on a substrate/metasurfaces

The preceding discussion suggests that dissolving a collection of gold NCs in water is a viable approach to create a radiation filter with a transmission window on the visible–near IR, with the window tuned to the desired frequency band and the resulting attenuation of the surrounding spectrum. However, given that the particles suspended in the fluid may change their average relative distances with time, or even deposit and aggregate at the bottom of a container under certain conditions, this approach may not be durable. An alternative is to lock the inter-particle distances and relative orientations by patterning them over a substrate, using lithographic techniques. The resulting system would be a structured metasurface composed by a dielectric substrate and a collection of metallic patterns (Figures 4a and 5a). This alternative has also other advantages: it can potentially produce results comparable to the solution with a much shorter optical length and the form factor may make it better suited for protecting static devices, such as circuit components, or even to be built-in as part of its encasing. But, for the filtering to be adequate, the proportion of surface area covered by the metal should be as large as possible, while keeping enough NC-NC distance as to minimize NC interaction.

In this section we present theoretical and experimental results showcasing this approach and discussing its limitations. With the discussed design we have aimed to use a collection of NCs with a wider range of sizes and different geometries. It will cover a different, wider spectral region and the target frequency for the transparency window will also be shifted into the IR region, around $\lambda = 1300\ nm$ (Figure 4b). We note that with colloidal nanorods it seems challenging to design metafluids for the infrared region $\lambda > 1000\ nm$ since the colloidal nanorods available for this spectral region become more polydisperse.[10] Simultaneously, the near and short-wavelength IR regions can be covered with lithographic structures and this is why we focus in this section on this type of nanocrystals.

In the metasurface case the design for the NCs geometries will be different, as the constraints are also different. First, if the sample is to function in an arbitrarily thin substrate, as opposed to submerged in an optically thick dielectric, we cannot rely on extra extinction beyond what is provided by the metal. Thus, there is a need to extend the collection of geometries to cover a broader spectral range.

An additional constraint comes from the need to maximize surface coverage by gold to maximize its surface density and thus the optical effect. Given that we need some buffer space around each structure to ensure that the interaction between them is weak, a clear path to maximize coverage is to minimize the number of independent structures. Of course, this limitation could be relaxed if we consider a multilayered design, which would increase the optical length traversed by light. However, the fabrication of such a system is non-trivial and we restricted this study to a single layer of metamaterial. So, to effectively satisfy these two constraints we should design a collection of structures with resonances over a wide range of wavelengths, while at the same time we keep the number of particles as low as possible. We approached this issue by using some structures shaped as crossed rods. These effectively contribute with more than one resonance, so they serve to increase the opacity both below and above the transparency window. Similar geometries have been already discussed in our previous work, in ref 9. The individual responses of the NCs involved can be seen in Figure 4b, as well as their sum, which depicts the result of ideally combining their effects without interaction. The cross-section sum is compared with the cross-section of the combined system (using the layout depicted in Figure 4a, to which we refer as dense layout in subsequent figures) under orthogonal light incidence. Their similitude shows that the optical distortion of the effect due to NC-NC interaction at that distance is weak.

The actual response of the system is also going to be impacted by the neighboring cells on the substrate. In what follows, the simulations for the metasurface describe periodic systems with a dielectric substrate (fused silica data from ref 27), and considering normal light incidence (Figure 5a).

Figure 5b shows the comparison of the experimental transmission profiles with theoretical results, obtained for layouts with the same cell periodicity and similar inter-NC distances. The overall agreement between experiment and theory is good, and we find that the transparency window effect is appreciable in the experimental sample. As expected, a more densely packed layout increases the global extinction of the sample and provides a better contrast. The sparser experimental sample (Figure 5b, black outline) was fabricated using as a model the layout proposed in ref 9, while the denser sample (Figure 5b, in red) uses a layout that increases gold surface coverage. Surface area covered by gold goes from ~4% to ~9% with this layout redesign. At the same time, interparticle distances decrease as well. We suggest the following metric to obtain a global measurement of particle proximity: if we define $d_m$ as the minimum distance that separates one NC from the closest neighboring NC, we can use its average over all structures, $\overline{d_m}$, to characterize the general interparticle separation. We compute values for $\overline{d_m}$ of 266 and 129 *nm* for the sparse and dense layouts, respectively. This decrease in interparticle distance has not introduced important changes into the optical profile of the planar metasurface.

Clearly, neither the window effect nor the overall extinction spectrum would resist an arbitrary reduction of interparticle distance. Now we will examine how an extreme scenario would compare with the layouts that we used. Figure 6 exhibits simulation results for a range of systems that differ in the level of structure packing and in the size of the periodic cell, while sharing the same relative abundance of geometries. Increasing the Au surface density decreases the overall metasurface transmission, but we also observe how, upon bringing the NCs together, the window feature is either strongly redshifted or destroyed altogether when the NCs are in contact. These compact and contiguous cases represent an extreme, however, as they involve a $\overline{d_m}$ of 5 *nm* in one case and zero in the other. Although it is notable that in the compact configuration we still observe a transparency window, its center is considerably shifted from the wavelength region intended by design. It hints nonetheless to the possibility of attempting intermediate configurations. Of course, if one is to aim for small interparticle distances, it would be reasonable to expect that some fraction of the NCs in the sample would be in contact with one or more of its neighbors. If that proportion is large enough, the window feature would be destroyed, as shown in the relevant simulation results in Figure 6 (blue). The system labeled as contiguous keeps the general structure organization displayed in the other systems, but with the NCs densely packed and in direct contact with the surrounding structures. In addition to the window disappearance, it also becomes difficult to distinguish features of the individual NC modes.

These two examples of strong near field interaction contrast with the two sparser layouts. In the same Figure 6 we have included the expected transmission for the fully independent collection of NCs (dashed curve in Figure 4b), as computed by using eqs 1 with $nL = 1/A$, being $A$ the surface area of the periodic cell of the dense layout. Thus, the two red curves correspond, respectively, to the fully periodic simulation of the dense layout (continuous) and the prediction for a non-interacting collection (dashed). So, from their similarity we can argue that the effects of NC-NC interactions in the periodic dense layout are relatively weak and are still a good representation of the original intended design.

Another useful perspective to highlight the relevance of the NC interaction comes from observing field maps of the excited system, which make apparent the spatial extent and geometry of the different resonant modes. In Figure 7a we show selected field enhancement ($FE = |\boldsymbol{E}_\omega(\boldsymbol{r})|/E_0$) maps for both the dense and compact system. These maps are representative of the rest of the spectrum in that they show how the field enhancement changes from being generated by single particle modes (as it is intended in the original design process) in the dense layout to become dominated by interparticle hot-spots in the compact layout. The description of the system goes from being fundamentally linear on the different single particle oscillation modes to being highly non-linear on them due to their strong interaction. Then, going one step further and considering the contiguous layout with all NC in contact, it might be more appropriate to consider the system as one single structure with an exceptionally complicated geometry.

**Experimental details for the lithographic metastructures.** In the experimental setup for the metastructure, double-side polished fused silica is selected as the substrate for the planar structures, owing to its well-known optical properties and high transparency in the spectral range of interest. Structures are realized on a 2 × 2 mm area using electron beam lithography technique followed by consecutive gold deposition by thermal evaporation. Optical transmission is measured using Fourier transform infrared spectroscopy system. Defined overall structure area was sufficient to measure optical transmission in the far-field without need for a microscope attachment.

## 5. Another realization of complex metafluid: Metastructure flakes in solution

We have seen two different approaches to the basic design concept, superposing individual particle resonances. One involves a colloidal solution of the NCs, while the other uses controlled nanostructures deposition on a substrate. We have focused part of the discussion on the deleterious effects of NC-NC interactions for the purposes of creating a stable transmission window. In a solution, we are in principle giving up control over the evolution of average NC distance over time. On the other hand, when we design and create a planar layout over a substrate we have control over the nanostructures position and orientation, but we are limited in the maximum optical path length that we can achieve.

A possible third approach that would combine advantages of these two would be the creation of a solution of gold metaflakes: individual substrate units containing a full layout of gold NCs (Figure 8a). One possible fabrication procedure would entail the creation of a full metasurface on an adequate substrate material, to then etch the substrate and break the metastructure into pieces. These would be transferred into some solvent. Such technique should also involve stabilization of the plasmonic NCs using special ligands that will prevent the structure from dissolving and aggregating. Some technologies for release of lithographic nanocrystal to a solution were recently reported. This mixed approach would allow control over NC-NC distances, avoiding global NC aggregation. On the other hand, a solution of randomly arranged flakes in a macroscopic solution (width $\sim 1\,cm$) increases the optical path traversed by light without need for considering the potentially more complicated fabrication process of a multilayer metasurface. This metaflake solution would also resistant to macroscopic stresses.

In Figure 8b we show the theoretical expectation for a simple transfer of previous metastructure results, integrated into a solution. We have taken simulation data for randomly oriented metaflakes in a liquid dielectric matrix and used eq 1 with $nL = 5 \cdot 10^8\,cm^{-2}$. Such plasmonic metaflakes dispersed in a liquid solvent should be based on thin dielectric substrates. For simplicity, our calculations were done for uniform dielectric environment, which can be achieved using an index-matched solvent. The geometries used in the simulation of the metaflakes are equivalent to the one described in the previous section, but the nanorod and crosses lengths have been slightly extended, thus the general redshift of the spectrum, including the transparency window, and the increased spectral coverage at large wavelengths. Another important factor for experiments with infrared plasmonic structures is absorption of a matrix, which can mask the plasmonic effects. For example, experiments with a transmission window in the interval 1-2μm can be done in a toluene solvent that does not absorb strongly in this spectral interval. Water is good matrix for the systems shown in Figures 1-3,[10] since it is transparent for the interval $\lambda < 1200\,nm$.

# Conclusions

We have discussed the design and implementation of systems that offer extinction profiles on a broad band of frequencies while preserving a narrow transparency window built within it. We have explored examples of systems with such a window in the visible-near IR spectrum, but the principles of design presented here should allow for a selection of different target frequency ranges. The different execution approaches described (NC solution, metasurface and metaflake solution) offer distinct benefits and limitations, so each of them may be adequate for different purposes. Throughout their presentation we have focused part of the discussion on the effect of NC-NC interaction as a hindrance to achieve the intended effect. We have explored theoretical scenarios in which the interaction does indeed destroy the transparency window effect, to better understand the limits of this method's applicability. We have seen that such limits, to the extent that they concern NC-NC interaction, provide ample room for realizable designs, and we have reported novel experimental results showing that with lithographically fabricated metasurfaces.

In summary, we have described a general design method, executed through different approaches, to create either isotropic or directional broad spectrum frequency selectors with high spectral precision, as well as discussed their limitations coming from the NC-NC interactions. These systems could find use in different applications, such as electromagnetic shielding for optoelectronics, spectral filters in both incoming and outgoing communication devices or as components in a multilayered energy harvesting system.

**Supporting Information**

This information includes detailed geometrical description of the metasurface layout, additional SEM images of the experimental system and details about the material data used in the simulations.

## AUTHOR INFORMATION

**Corresponding Authors:**

*E-mails: lvbesteiro@gmail.com and govorov@ohiou.edu.

Notes: The authors declare no competing financial interest.

## ACKNOWLEDGMENTS
L.V.B. and A.O.G. were supported by the U.S. Army Research Office under Grant Number W911NF-12-1-0407 and by the Volkswagen Foundation (Germany). K.G. and H.V.D gratefully acknowledge the financial support from Singapore National Research Foundation under the program of NRF-NRFI2016-08 and the Science and Engineering Research Council, Agency for Science, Technology and Research (A*STAR) of Singapore; EU-FP7 Nanophotonics4Energy NoE; and TUBITAK EEEAG 115E679. H.V.D. acknowledges support from ESF-EURYI and TUBAGEBIP. K.G. acknowledges support from TUBITAK BIDEB 2211 program.

# Figures

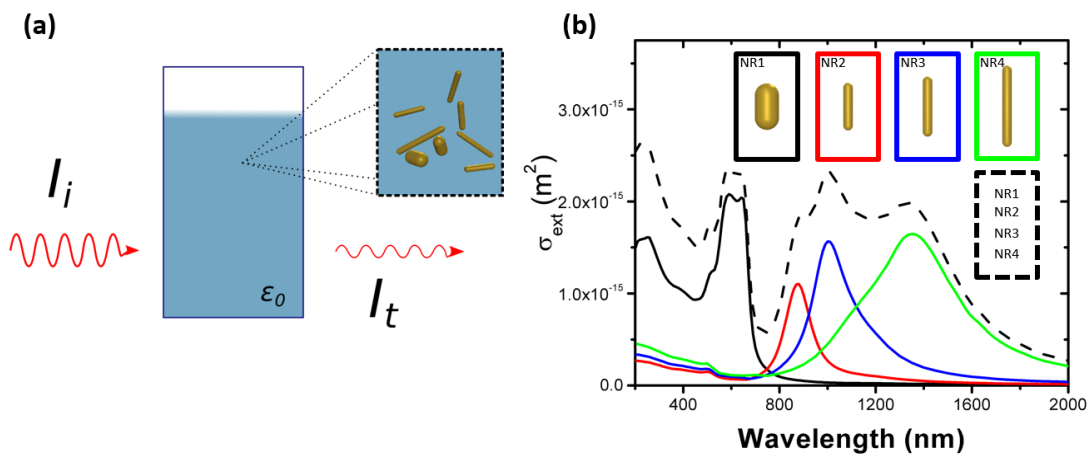

Figure 1: (a) Illustration of a dispersion of gold colloidal NCs in a solvent matrix. The solution will be opaque to a wide range of light frequencies, but transparent in a narrow spectral range. (b) Extinction cross sections for the different rod geometries and their sum. The particle sizes are selected so that their extinction profiles create the transparency window around $\lambda = 740\ nm$. Each extinction cross-section curve is obtained by averaging over incident angle and polarization.

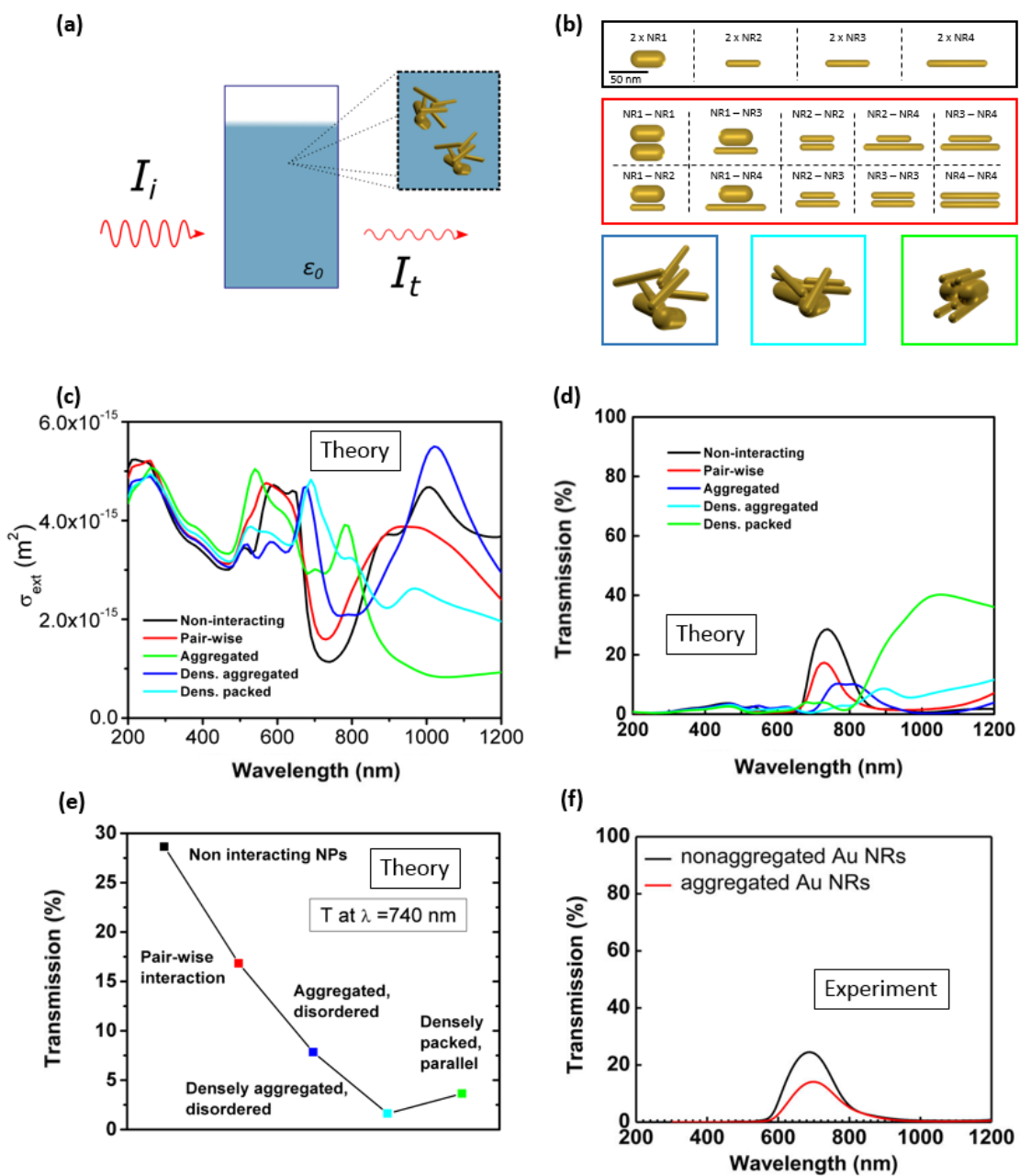

Figure 2: (a) Illustration of the solution, populated by cells constructed to probe different levels of NC packing. (b) Systems described by the legends in panels (c) to (e). (c) Extinction cross sections for each of the individual cells considered. (d) Theoretical predictions for the transmission, obtained from the extinction cross sections of panel (c). Results are obtained for a cell concentration of $n = 1.1 \cdot 10^{11}$ $units/cm^3$ and $L = 1\,cm$. (e) Data points taken at $\lambda = 740\,nm$ from panel (d). This narrows the transparency window to its nominal spectral center to facilitate the observation of its collapse. (f) Transmission from experimental samples at different levels of particle aggregation, reprinted with permission from ref 10. Copyright 2016 American Chemical Society.

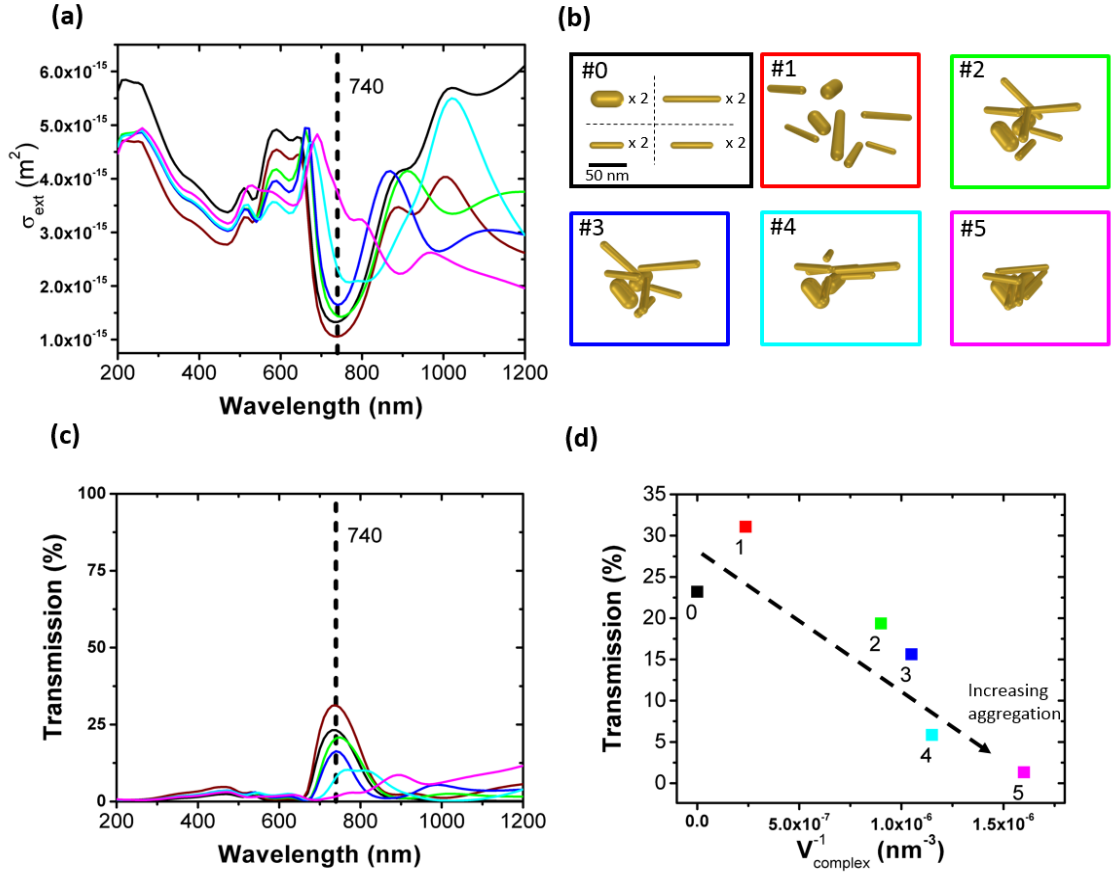

Figure 3: (a) Extinction cross sections of isolated cells containing systems at different levels of aggregation. (b) Representation of such systems. Their colored frames serve as a legend for the other panels in the figure. They are arranged in order of increasing compactness. (c) Transmission obtained from the data in panel (a), using $n = 1.1 \cdot 10^{11}\ units/cm^3$ and $L = 1\ cm$. (d) Data points extracted from panel (c) at $\lambda = 740\ nm$. The magnitude $V_{complex}$ is the volume of a box containing the eight NCs in a unit cell. Superimposed to a global tendency of increasing opacity with NC aggregation we see important deviations due to the specific NC-NC interactions.

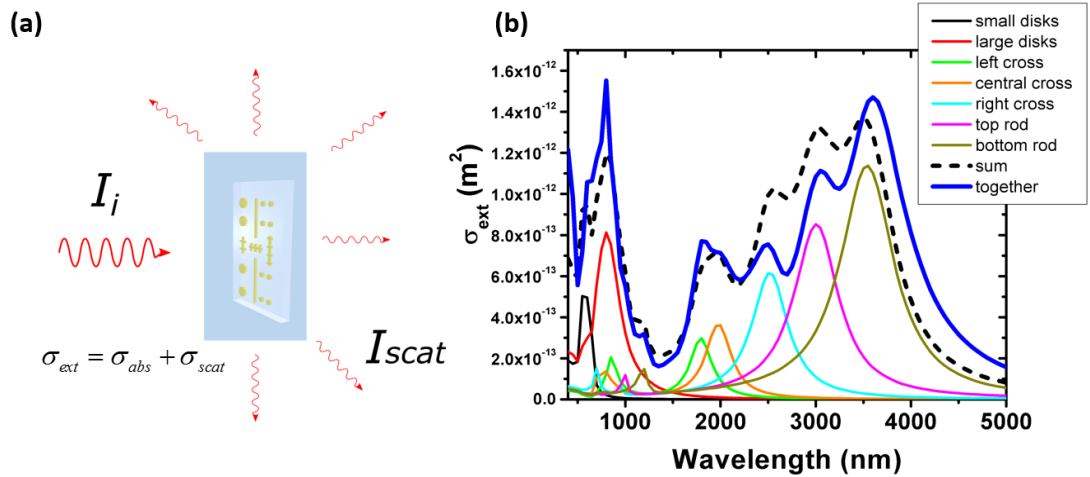

Figure 4: (a) Illustration of a single instance of a metasurface layout over substrate, isolated and surrounded by air. (b) Extinction cross-section of the different gold NCs used in the layout, under normal illumination (curves for disks' cross-sections represent the total extinction for all disk instances of that size in the layout, not for individual ones). The dashed curve is the sum of the individual non-interacting particles. The continuous blue curve was obtained with the NCs arranged in a dense layout, depicted in panel (a) and in Figures 5-7.

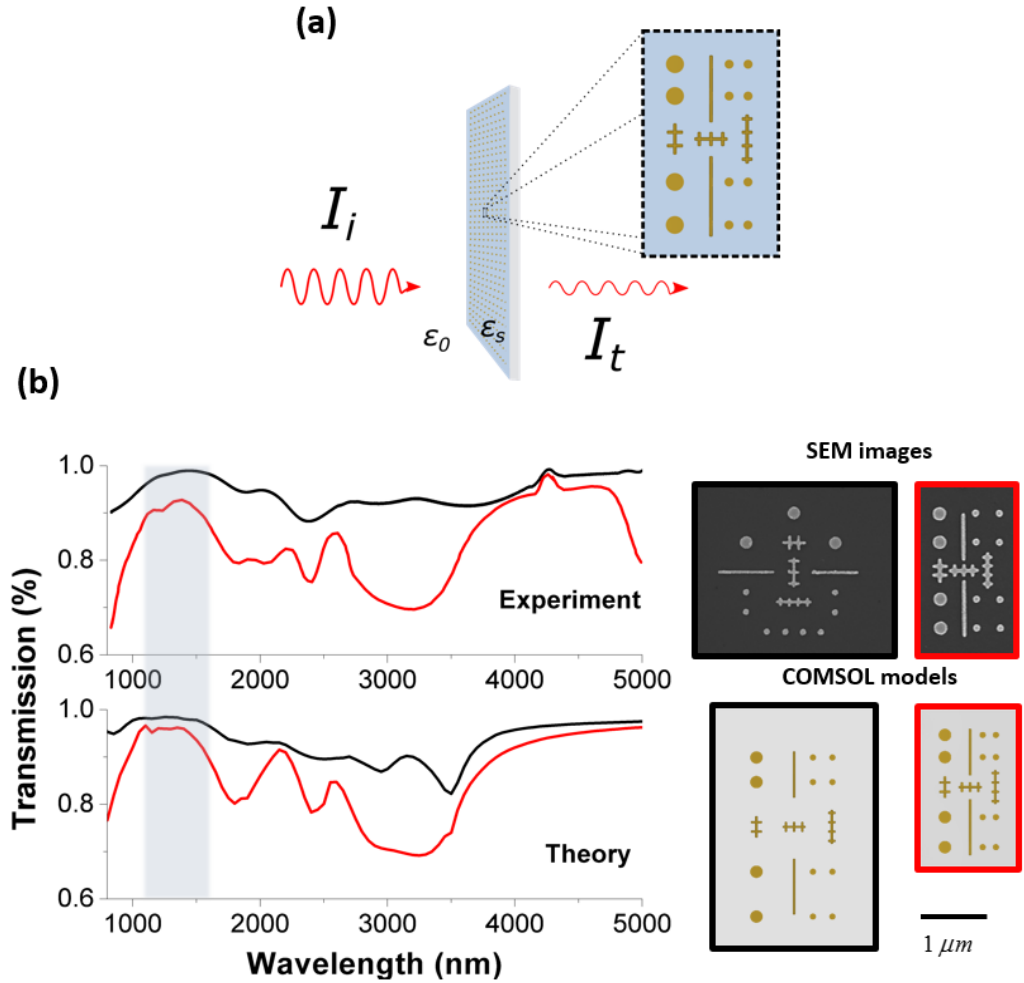

Figure 5: (a) Illustration of a metasurface composed of a periodic layout of gold NCs lithographically deposited over glass substrate. (b) Comparison of metasurface transmission between experiment (top panel) and simulations (bottom panel). There are two curves in each panel, corresponding to the systems depicted in the images on the right. Black curves show data from sparse layouts, while the red curves show results for dense layouts. A light blue bar in the plots signals the transparency window. The two top insets on the right are SEM images of the experimental samples, while the models below depict the simulated systems.

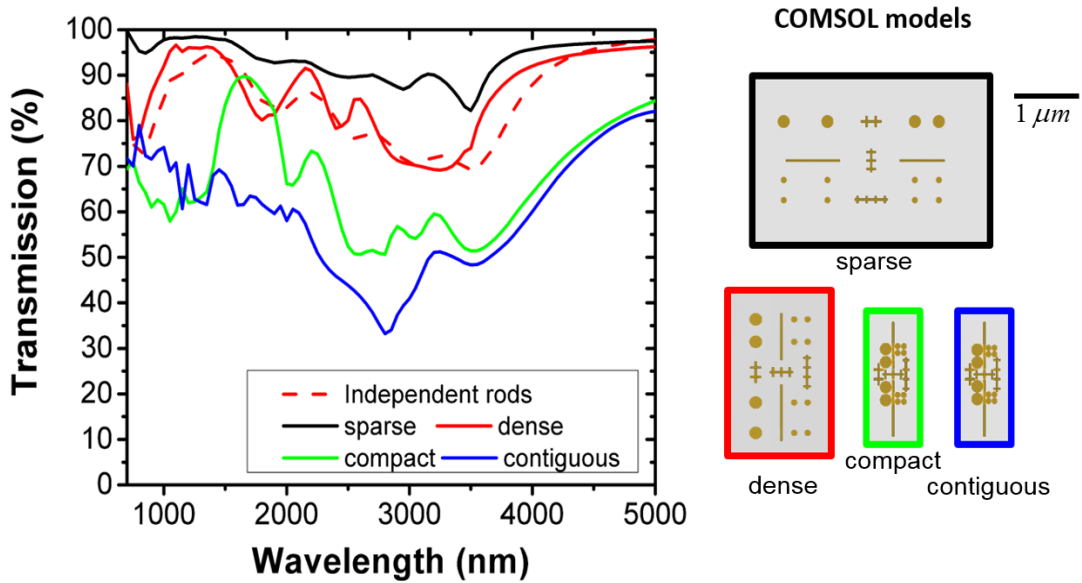

Figure 6: The continuous curves are a collection of simulated results for periodic systems with differently packed layouts: sparse, dense, compact (5 $nm$ interparticle distance) and contiguous. The red dashed line is obtained from fully independent NC cross section results (Figure 4b, dashed curve) in a system with $nL = 1/A$. Panels on the right are models of the layouts used for these simulations.

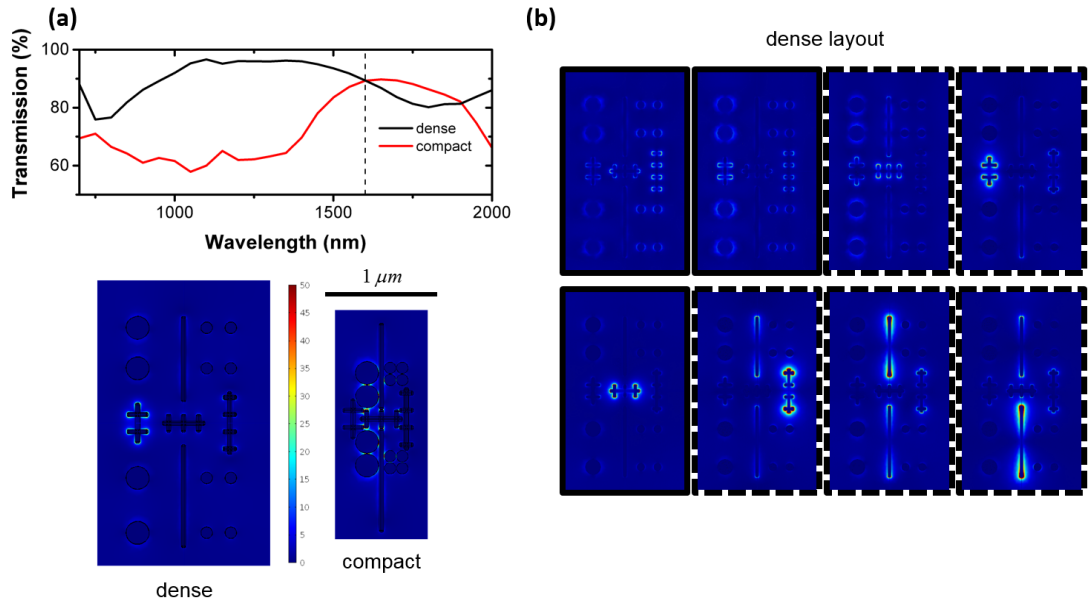

Figure 7: (a) Detail of the transmission spectrum of dense and compact layouts. The vertical line marks the wavelength used to obtain the data on the images below. These are Field Enhancement maps for the dense and compact layouts, respectively, under light linearly polarized along the long axis of the system. (b) Some representative resonances in the dense layout, shown as $FE$ maps. Images with continuous frame are obtained with long axis linearly polarized light, while the ones with dashed frame used short axis polarization. Their free space wavelengths, from left to right and top to bottom are: 0.7, 0.75, 0.8, 1.8, 1.95, 2.5, 2.85 and 3.5 $\mu m$. For clarity, all images share the same color range scale (color bar in panel (a)), but the maximum $FE$ in several of them is above the maximum depicted value.

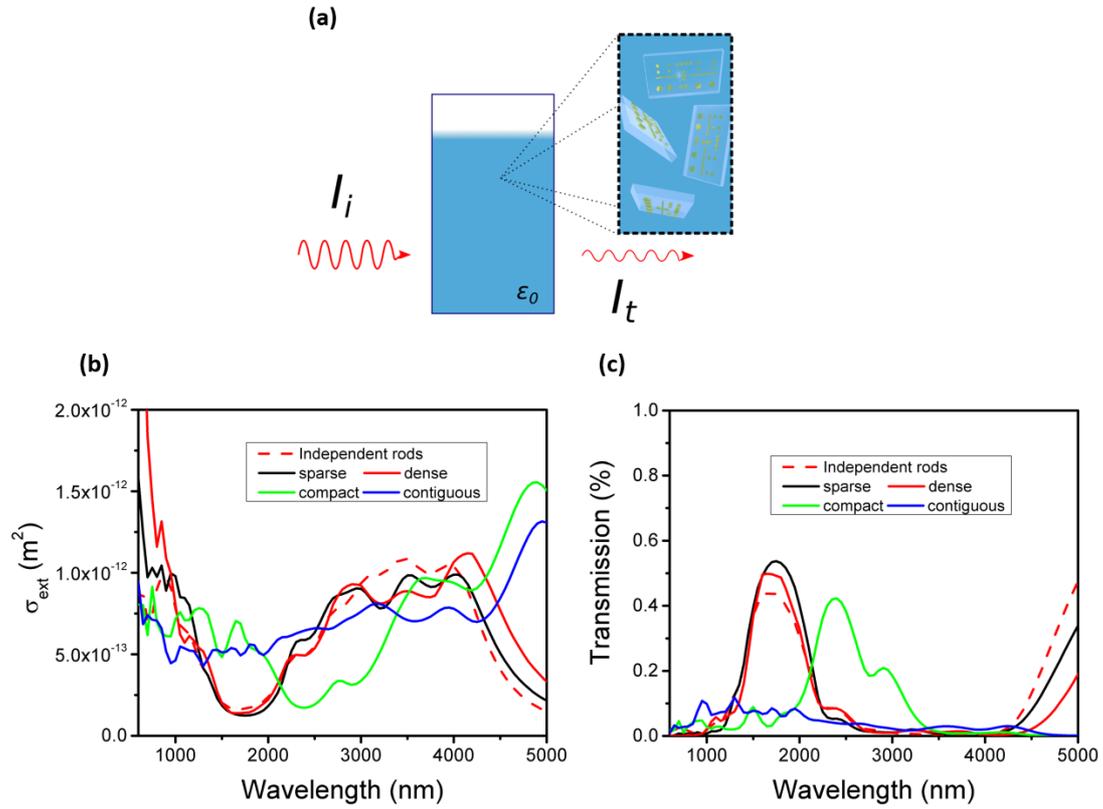

Figure 8: (a) Illustration of a system of metaflakes in a solution with homogeneous dielectric environment. Each individual unit is composed of an ordered collection of NCs. (b) and (c) Extinction cross sections and transmission data, respectively, for metaflake solutions, as obtained with numerical methods. We show results for non-interacting NCs and four flake layouts with decreasing NC separation: sparse, dense, compact and fully contiguous. Cross sections are averaged over the combination of three incidence directions and two light polarizations. Transmission calculated using $nL = 5 \cdot 10^8 \, cm^{-2}$.

# Supplementary Information

# Simple and Complex Metafluids and Metastructures with Sharp Spectral Features in a Broad Extinction Spectrum: Particle-Particle Interactions and Testing the Limits of the Beer-Lambert Law

Lucas V. Besteiro[1*], Kivanc Gungor[2], Hilmi Volkan Demir[2,3], Alexander O. Govorov[1,4*]

[1] Department of Physics and Astronomy, Ohio University, Athens, Ohio 45701, United States

[2] Department of Electrical and Electronics Engineering, Department of Physics, UNAM-Institute of Materials Science and Nanotechnology, Bilkent University, Ankara 06800, Turkey

[3] Luminous! Center of Excellence for Semiconductor Lighting and Displays, School of Electrical and Electronic Engineering, Division of Physics and Applied Physics, School of Physical and Mathematical Sciences Nanyang Technological University, Singapore 637371, Singapore

[4] Institute of Fundamental and Frontier Sciences, University of Electronic Science and Technology of China, Chengdu 610054, People's Republic of China

**Geometric detail of metastructure layout.**

The systems discussed in the main text include a two-dimensional patterned metasurface. Here we describe geometric details of the simulated system (Figure S1) and showcase images of the experimental layout at different scales (Figure S2) to offer a better perspective of the periodicity of the system. In Figure 8, we used slightly different geometrical parameters, as compared to the metastructure parameters given in Figure S1. Correspondingly, the position of the transmission window in Figure 8 is slightly different to that in Figure 4.

In Figures 5 and 6, the lateral periods of the metasurface used in the simulations were taken in the following way (width x height, both in nm):
Sparse: 2500x3800.
Dense: 1500x2500.
Compact: 800x2000.
Contiguous: 800x2000.
The above periods in our simulation were chosen close to those in the experimental structures. In particular, the experimental periods in the metasurfaces shown in Figure 5 had the unit cells: 3000x3000 (sparse) and 1500x2500 (dense).

## Material data in the simulations.

### Gold data in Au NRs in solution:

Our simulations concerning metafluids used a broadened version of gold experimental permittivity data. We used Drude model to broaden the experimental data, in the following way:

$$\varepsilon_{br} = \varepsilon_{exp} - \varepsilon_{Drude} + \varepsilon_{Drude,br}$$

$$\varepsilon_{Drude} = \varepsilon_{\infty} - \frac{\omega_P^2}{\omega(\omega + i\gamma_P)}$$

$$\varepsilon_{Drude,br} = \varepsilon_{\infty} - \frac{\omega_P^2}{\omega(\omega + 3i\gamma_P)}$$

where $\varepsilon_{exp}$ is the permittivity data obtained from the literature, $\omega_P$ and $\gamma_P$ are the plasma frequency and damping rate, respectively. For gold, these parameters take the values $\omega_P = 8.9\ eV$ and $\gamma_P = 0.076\ eV$.

### Matrix data in non-periodic planar structures:

Extinction data for individual NCs and the metastructure complex, shown in Figure 4b in the main text, was obtained by simulating isolated, non-periodic systems. The permittivity of the background matrix was selected so that it approached the environment of the metastructure deposited over the fused silica substrate. We used a constant matrix permittivity of $\varepsilon_0 = 1.5$, averaged from the values of $\varepsilon_{f.silica} \approx 2$ and $\varepsilon_{air} = 1$.

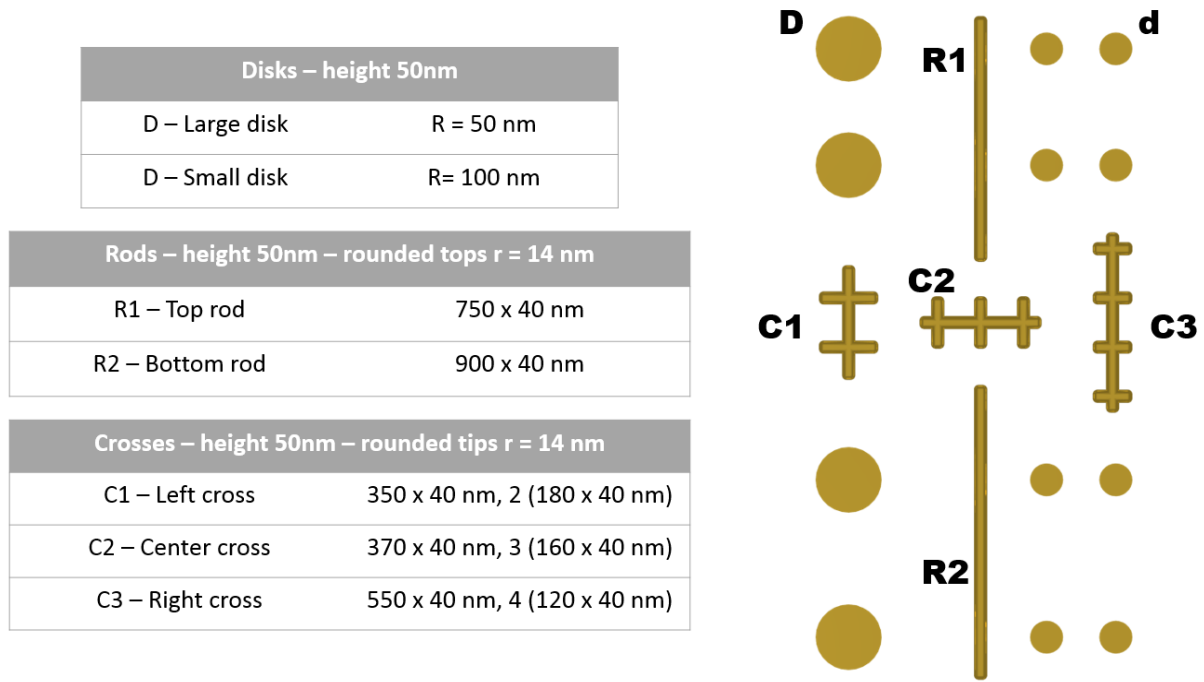

| Disks – height 50nm | |
|---|---|
| D – Large disk | R = 50 nm |
| D – Small disk | R= 100 nm |

| Rods – height 50nm – rounded tops r = 14 nm | |
|---|---|
| R1 – Top rod | 750 x 40 nm |
| R2 – Bottom rod | 900 x 40 nm |

| Crosses – height 50nm – rounded tips r = 14 nm | |
|---|---|
| C1 – Left cross | 350 x 40 nm, 2 (180 x 40 nm) |
| C2 – Center cross | 370 x 40 nm, 3 (160 x 40 nm) |
| C3 – Right cross | 550 x 40 nm, 4 (120 x 40 nm) |

Fig S1. Geometric details of the NCs used in the simulations of the metasurface. Their relative positions are representative of the layout described as "dense" in the main text.

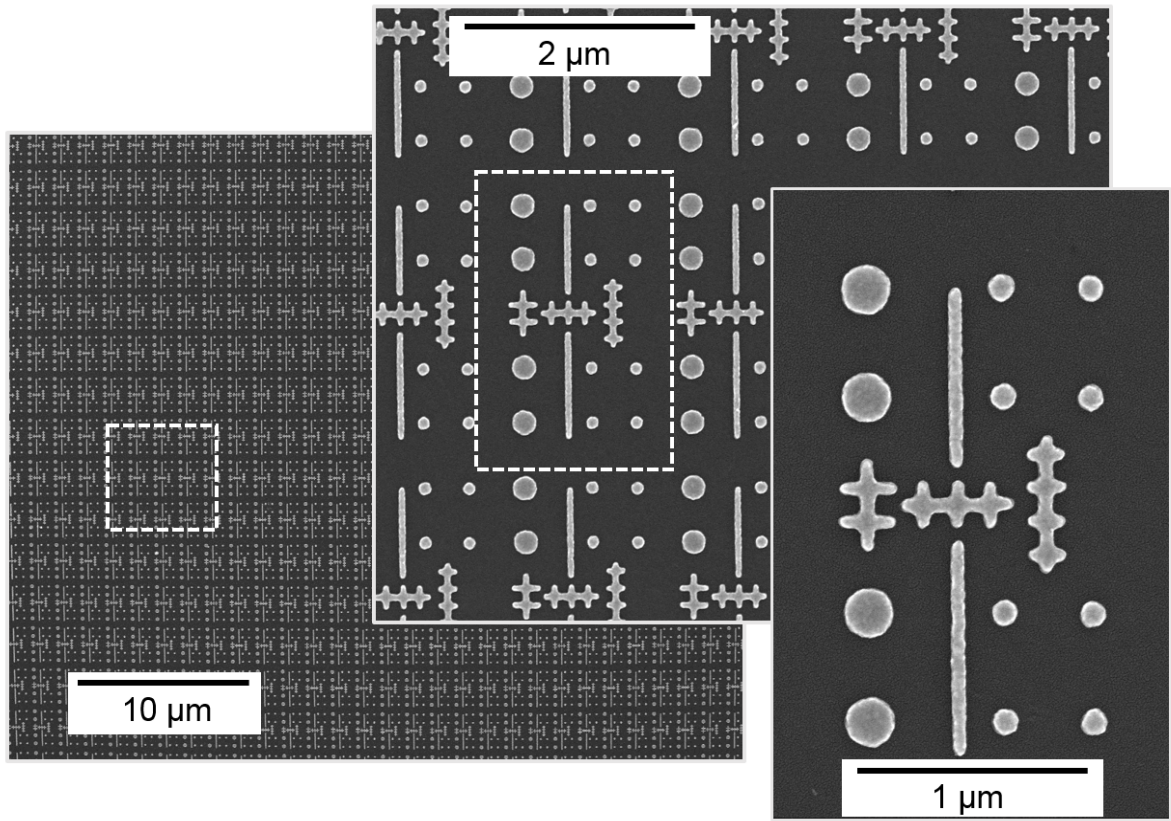

Fig S2. SEM images at different scales of one of the metasurfaces described in the main text. This metasurface was regarded as a structure with the dense layout (Figure 5). The images in this figure illustrate the replicated pattern of gold nanostructures over the fused silica substrate. Optical transmission for this structure is shown in Figure 5 in the main text.